\documentclass{aa} 

\usepackage{graphicx}
\usepackage{txfonts}
\usepackage{longtable}
\usepackage{amsmath}
\usepackage{graphicx}
\usepackage{float}
\usepackage{multirow}
\usepackage{subfigure}
\usepackage{rotfloat}
\usepackage{xcolor}
\usepackage[]{hyperref}
\usepackage{breakurl}

\usepackage{array}
\usepackage{lineno}

\usepackage{booktabs}
\usepackage{lscape}
\usepackage{subfigure}
\begin{document}

\title{Self-organized critical characteristics of teraelectronvolt photons from GRB 221009A}

\author{Wen-Long~Zhang\inst{1,2}
\and Shuang-Xi~Yi\inst{1}
\and Yuan-Chuan~Zou\inst{3}
\and Fa-Yin~Wang\inst{4}
\and Cheng-Kui~Li\inst{2}
\and Sheng-Lun~Xie\inst{5}
}

\institute{$^1$ School of Physics and Physical Engineering, Qufu Normal University, Qufu 273165, China\\
\email{yisx2015@qfnu.edu.cn}\\
$^2$ Key Laboratory of Particle Astrophysics, Institute of High Energy Physics, Chinese Academy of Sciences, Beijing 100049, China\\
$^3$ School of Physics, Huazhong University of Science and Technology, Wuhan 430074, China\\
\email{zouyc@hust.edu.cn}\\
$^4$ School of Astronomy and Space Science, Nanjing University, Nanjing 210023, China\\
$^5$ Institute of Astrophysics, Central China Normal University, Wuhan 430079, China}

\markboth{Self-organized critical characteristics of teraelectronvolt photons from GRB 221009A}{}



\abstract{The very high-energy afterglow in GRB 221009A, known as the “brightest of all time” (BOAT), has been thoroughly analyzed in previous studies. In this paper, we conducted a statistical analysis of the waiting time behavior of 172 TeV photons from the BOAT observed by LHAASO-KM2A. The following results were obtained: 
(I) The waiting time distribution (WTD) of these photons deviates from the exponential distribution. 
(II) The behavior of these photons exhibits characteristics resembling those of a self-organized critical system, such as a power-law distribution and scale-invariance features in the WTD. The power-law distribution of waiting times is consistent with the prediction of a nonstationary process.
(III) The relationship between the power-law slopes of the WTD and the scale-invariant characteristics of the Tsallis q-Gaussian distribution deviates from existing theory. We suggest that this deviation is due to the photons not being completely independent of each other. 
In summary, the power-law and scale-free characteristics observed in these photons imply a self-organized critical process in the generation of teraelectronvolt photons from GRB 221009A. Based on other relevant research, we propose that the involvement of a partially magnetically dominated component and the continuous energy injection from the central engine can lead to deviations in the generation of teraelectronvolt afterglow from the simple external shock-dominated process, thereby exhibiting the self-organized critical characteristics mentioned above.}

\authorrunning{Zhang et al.}
\titlerunning{Self-organized critical characteristics of teraelectronvolt photons from GRB 221009A}
\maketitle

\section{Introduction}
Gamma-ray bursts (GRBs) are sudden high-energy explosions in the cosmos, releasing a large amount of gamma-ray photons. Currently, there is no evidence of their occurrence within the Milky Way. Gamma-ray bursts mainly consist of two primary phases: prompt and multiwavelength afterglow. The prompt phase is typically observed first when the gamma photon flux exceeds the threshold of high-energy photon monitoring satellites. Subsequently, other observation satellites are redirected to continuously monitor the potential multiwavelength afterglow following the prompt phase. Multiwavelength afterglow radiation spans a broad electromagnetic spectrum, including X-rays, radio, and optical bands, making it an indispensable tool for comprehensive GRB research.

The use of imaging atmospheric Cherenkov telescopes has greatly broadened the energy limits of our observations of GRBs, allowing us to see the ultrahigh-energy radiation components from the universe, including the very high-energy (VHE) afterglow of GRBs \citep{2019Natur.575..455M,2019Natur.575..459M,2019Natur.575..464A,2021Sci...372.1081H}. In particular, the Large High Altitude Air-shower Observatory (LHAASO) has yielded numerous surprises, including the detection of the VHE afterglow of the “brightest of all time” (BOAT) GRB 221009A \citep{2023ApJ...946L..31B,2023Sci...380.1390L,2023SciA....9J2778C,2023PhRvL.131y1001G} and the identification of 
the PeVatron \citep{2024SciBu..69..449L}. The emission of teraelectronvolt photons from the BOAT has been the subject of several studies, which have looked into such matters as the jet Lorentz factor \citep{2023ApJ...956L..38G}, Lorentz invariance violation \citep{2023JCAP...10..061L,2024arXiv240206009T}, megaelectronvolt-teraelectronvolt annihilation \citep{2024ApJ...961L...6G}, and dark matter candidate axion-like particles \citep{2022arXiv221102010C,2024JCAP...01..026G}, while also providing a fresh perspective on our cognition. As is mentioned by \cite{2024ApJ...961L...6G}, some researchers have suggested that these teraelectronvolt photons may originate from the prompt emission. Furthermore, according to \cite{2023arXiv231011821W} and \cite{2024arXiv240403229Z}, the teraelectronvolt emission from the BOAT may reflect signatures of the prompt megaelectronvolt emission. This leads us to speculate about whether it is possible that each teraelectronvolt photon is generated by a discrete event, and, if this is indeed the case, what connections exist between these events.

The concept of self-organized criticality (SOC) has been widely observed in various natural systems \citep{1988PhRvA..38..364B}. It is also known as the ``avalanche effect.''
The primary manifestation of SOC systems is the power-law distribution of statistical parameters such as time and energy for multiple events, as well as the characteristic of scale invariance \citep{2017ChPhC..41f5104C,2021ApJ...920..153W,2023PhRvR...5a3019W,2023ApJ...959..109P,2024ApJ...968...40G}.
Systems ranging from sand piles to celestial bodies such as the Earth \citep{1989JGR....9415635B,2007PhRvE..75e5101C}, the Sun \citep{2000ApJ...536L.109W,2003SoPh..214..361W,2013NatPh...9..465W,2023MNRAS.518.3959P}, black holes \citep{2017MNRAS.470.1101W}, soft gamma-ray repeaters \citep{1996Natur.382..518C,2017ChPhC..41f5104C,2017JCAP...03..023W,Cheng2020,2023RAA....23k5013Z}, some active galactic nuclei \citep{1984ARA&A..22..471R,2018ApJ...864..164Y}, and high-mass X-ray binaries \citep{2019MNRAS.487..420S,2022RAA....22f5012Z} have exhibited characteristics of self-organized critical systems in previous studies. Even in studies involving the prompt emission \citep{2021FrPhy..1614501L,2023ApJS..265...56L,2023ApJ...955L..34L,2024ApJ...965...72M} and the X-ray or optical afterglows \citep{2013NatPh...9..465W,2016ApJS..224...20Y,2017ApJ...844...79Y} of GRBs, as well as research on repeating fast radio bursts \citep{2023ApJ...949L..33W,2023arXiv231212978W,Wu2024}, similar characteristics have been identified.

In the multiband emission of GRBs, the VHE radiation often displays relatively simple variability shapes, posing challenges for identifying SOC features using conventional methods. Fortunately, we have acquired teraelectronvolt photon data detected by LHAASO and reconstructed by \cite{2023SciA....9J2778C}. Section 2 presents the data source and the various analysis methods utilized. The findings from the data analysis are thoroughly discussed and analyzed in detail in section 3. Finally, our conclusions are presented in Section 4.

\section{Data and statistical analyses}
On October 9, 2022, at 13:17:00.00 (denoted as $T_0$ hereafter), the Fermi Gamma-ray Burst Monitor (GBM) was triggered by an extraordinarily bright GRB, GRB 221009A \citep{2022GCN.32636....1V}. At the same time, it also triggered several high-energy satellites such as Swift-BAT, Insight-HXMT, GECAM-C, and Konus-Wind \citep{2022GCN.32635....1K,2023arXiv230301203A,2022GCN.32668....1F}. Subsequent optical observations determined the redshift of the GRB to be z =0.151 \citep{2022GCN.32648....1D}. LHAASO observed this GRB and detected more than 5000 VHE photons up to approximately 18 TeV, as is reported by \cite{2022GCN.32677....1H}.
We collected data on the arrival times of 172 TeV photons from GRB 221009A detected by LHAASO-KM2A, as is reported by \cite{2023SciA....9J2778C}.

\subsection{Waiting time distribution}
The waiting time distribution (WTD) has become an important tool for us due to the unique nature of the obtained samples. In this section, we introduce the arrival time intervals of adjacent teraelectronvolt photons as the waiting time, which is defined as \(T_{\text{waiting}}=T_{i+1}-T_{i}\), where $T_{i+1}$ and $T_{i}$ are the arrival times for the ($i + 1$)th and $i$th photon. We made two samples: one consisting of all 172 TeV photons, referred to as Sample \uppercase\expandafter{\romannumeral1}, and another one, referred to as Sample \uppercase\expandafter{\romannumeral2}, consisting of 143 TeV photons that arrived during the main emission time range of 230 s to 900 s (according to \citep{2023SciA....9J2778C}), in order to take out possible background.

Subsequently, we employed the exponential function, and the threshold power-law function to fit the distribution of waiting time. This was done to verify whether the overall process is generated by a completely random Poisson process and whether the distribution exhibits a power-law form. We then utilized the Tsallis q-Gaussian distribution \citep{1998PhyA..261..534T} to investigate its scale invariance characteristics. In this work, the python module $emcee$\footnote{https://pypi.org/project/emcee/} was utilized to fit the data and get the confidence intervals of the parameters with the Monte Carlo Markov chain method.

\subsubsection{Exponential function}
We first chose to fit the data distribution using an exponential function, which is expressed as follows:

\begin{equation}
f(x) \propto e^{-kx}
\label{eq_exponential}
,\end{equation}
where $k$ and the amplitude are free parameters for this function that need to be determined through fitting.

\subsubsection{Threshold power-law function}

In general, the differential distribution can be described by a threshold power-law distribution (also called a generalized Pareto type II distribution) with the following equation \citep{Arnold2008,2015ApJ...814...19A}:

\begin{equation}
\frac{dN}{dx}\propto (x+x_{0})^{-\alpha_{x}} 
\label{eq_diff}
.\end{equation}

The cumulative distribution can be expressed as the integral of the number of events exceeding a given value, x. Therefore, the cumulative distribution function corresponding to Equation (\ref{eq_diff}) can be expressed as ($\alpha_x\neq1$):

\begin{equation}
N_{\mathrm{cum}} (>x) = A+B\times(x+x_{0})^{1-\alpha_x},
\label{eq_cum}
\end{equation}

where $x_0$ is a constant by considering the thresholded effects (e.g., incomplete sampling below $x_0$, background contamination), $\alpha_x$ is the power-law slope of the distribution, $N_{\rm env}$ refers to the total number of events. The uncertainty of the cumulative distribution in a given bin, $i$, was approximately calculated as $\sigma_{\mathrm{cum}, i}=\sqrt{N_{\mathrm{cum,i}}-N_{\mathrm{cum,i+1}}}$ \citep{2019ApJ...880..105A}, where $N_{cum,i}$ and $N_{cum,i+1}$ are the number of events in the $i$th and $i+1$th bin, respectively.

The standard reduced chi-square ($\chi^{2}$) goodness was used to identify the best fit. The $\chi$ can
be written as
\begin{equation}
\chi_{cum}=\sqrt{\frac{1}{(n_x-n_{par})}\sum_{i=1}^{n_x}\frac{\left[N_{cum,th}(x_i)-N_{cum,obs}(x_i)\right]^2}{\sigma_{cum,i}^2}}
\label{}
\end{equation}
for the cumulative distribution function \citep{2019ApJ...880..105A}, where $n_{\mathrm{x}}\:$ is the number of logarithmic bins, $n_{\mathrm{par}}\:$ the number of the free parameters, $N_{\mathrm{cum,obs}}(x_{i})\:$ the observed values, and $N_{\mathrm{cum,th}}(x_{i})\:$ the corresponding theoretical values for the cumulative distribution, respectively. 

\begin{figure*}[http]
\includegraphics[width=0.5\textwidth, angle=0]{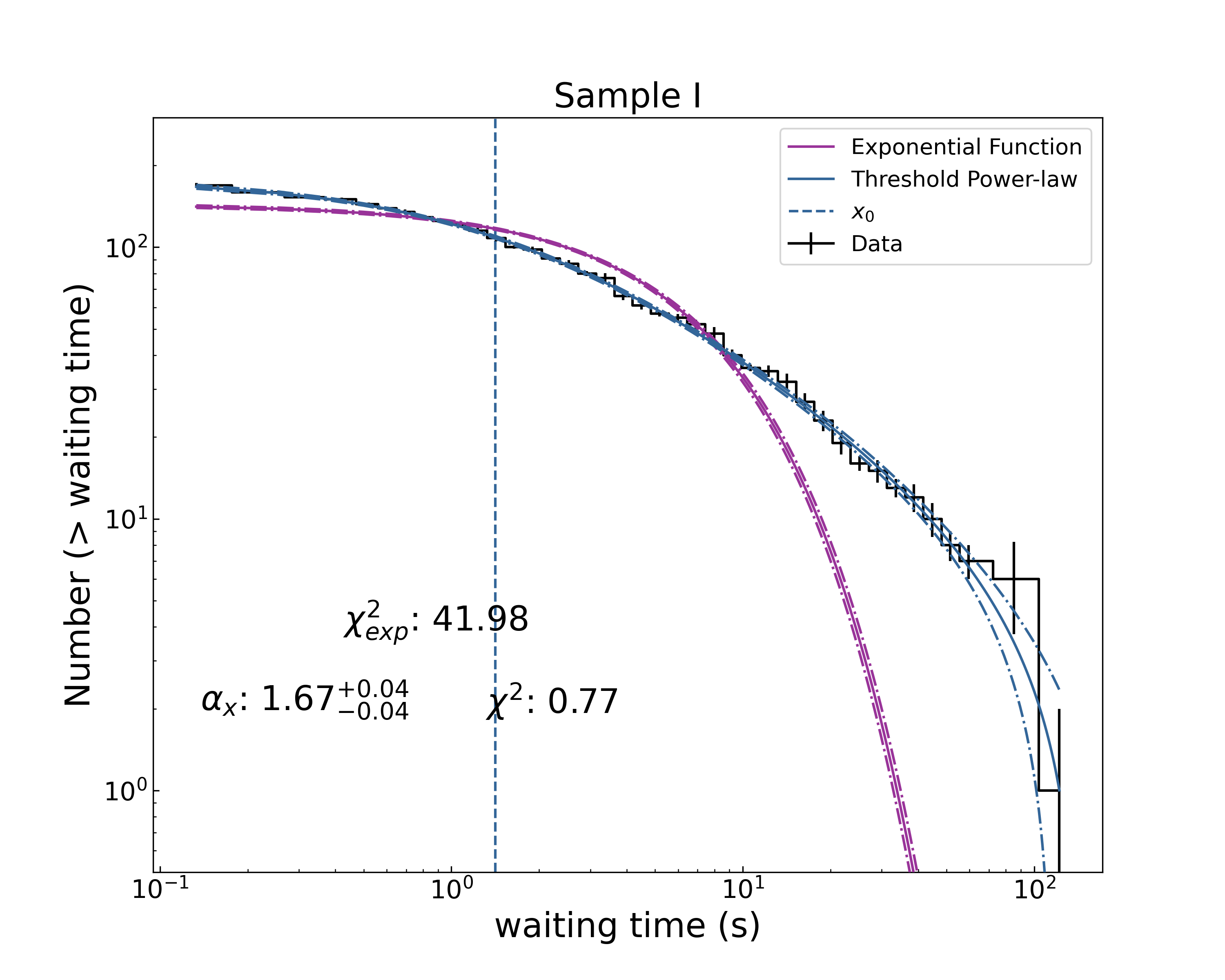}
\includegraphics[width=0.5\textwidth, angle=0]{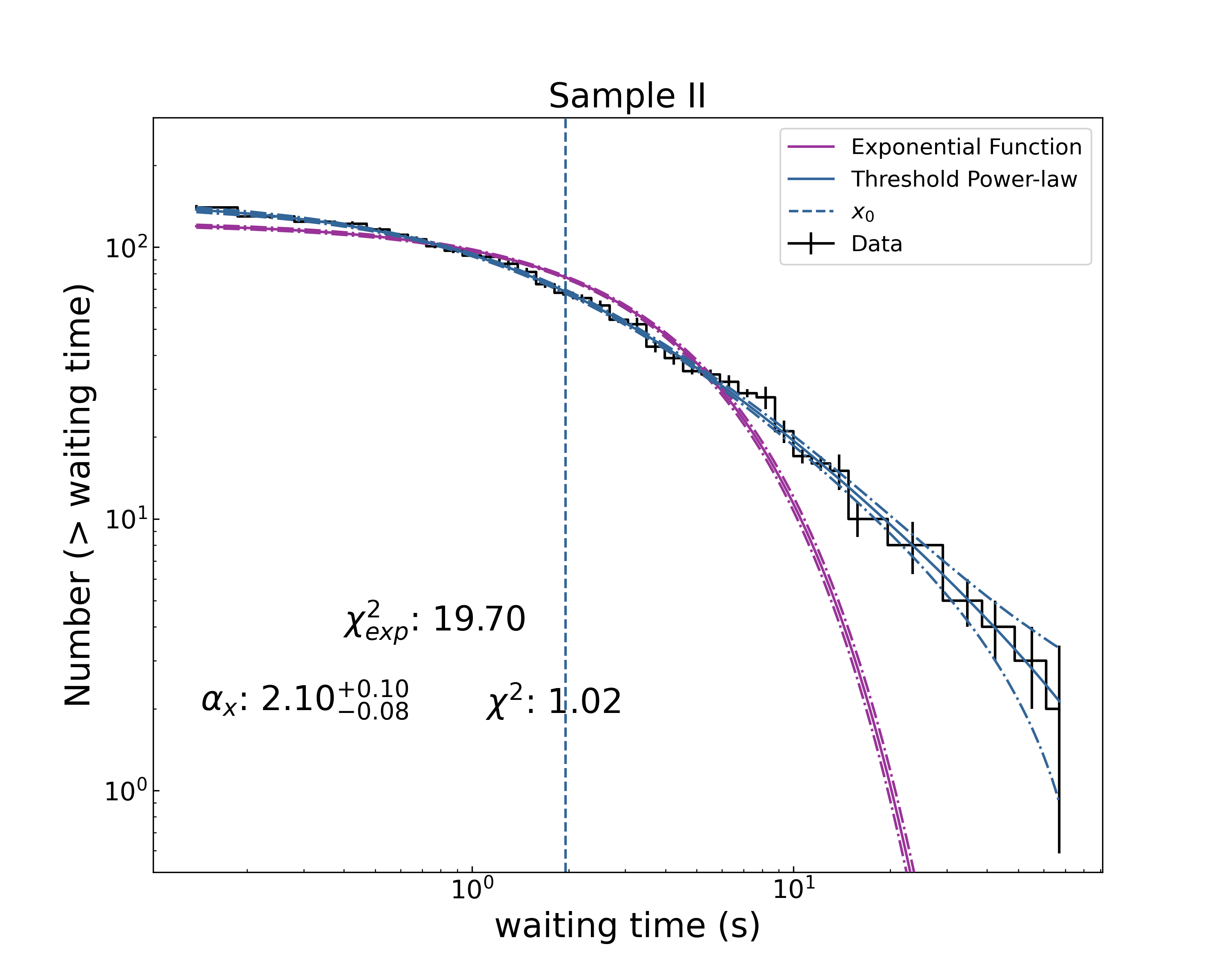}\\
\caption{Waiting time distributions and fitting results. The two panels  depict the cumulative distributions of waiting time for Sample \uppercase\expandafter{\romannumeral1} and Sample \uppercase\expandafter{\romannumeral2}, which have been fit with an exponential (Eq.~\ref{eq_exponential}) and a threshold power law (Eq.~\ref{eq_cum}). The solid lines represent the best-fitting results for each function, while the regions between the two dash-dotted lines indicate the 95$\%$ confidence level for each. Additionally, the dashed lines correspond to the thresholded values, $x_0$, for the threshold power law.}
\label{fit-wtd}
\end{figure*}

\subsection{Tsallis q-Gaussian distribution}

\begin{figure*}
\centering
\includegraphics[width=0.5\textwidth, angle=0]{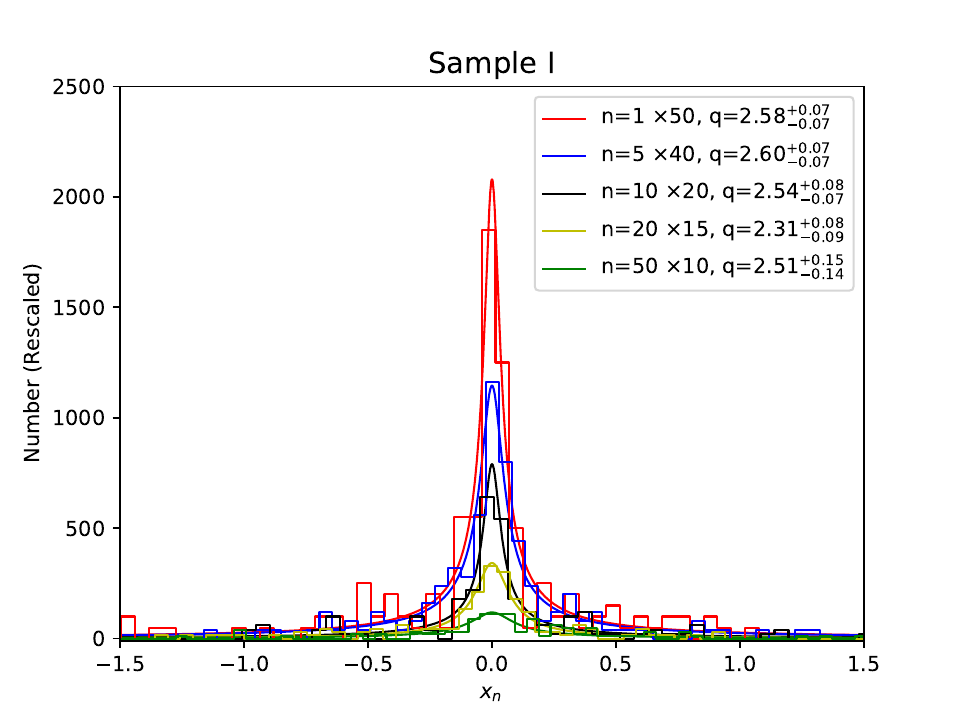}\includegraphics[width=0.5\textwidth, angle=0]{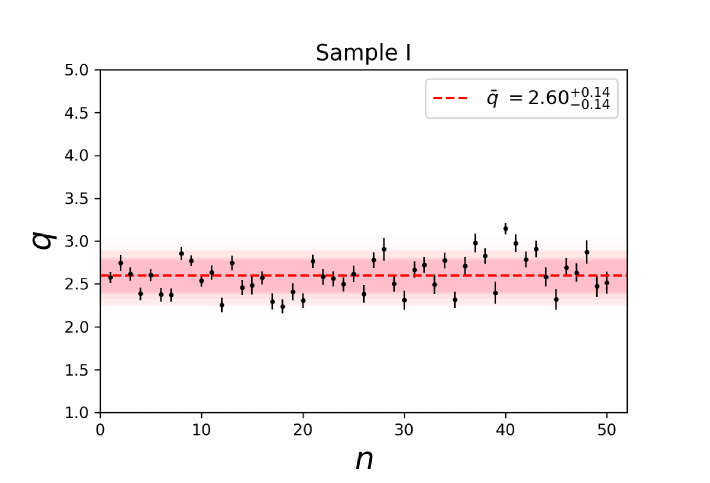}\\
\includegraphics[width=0.5\textwidth, angle=0]{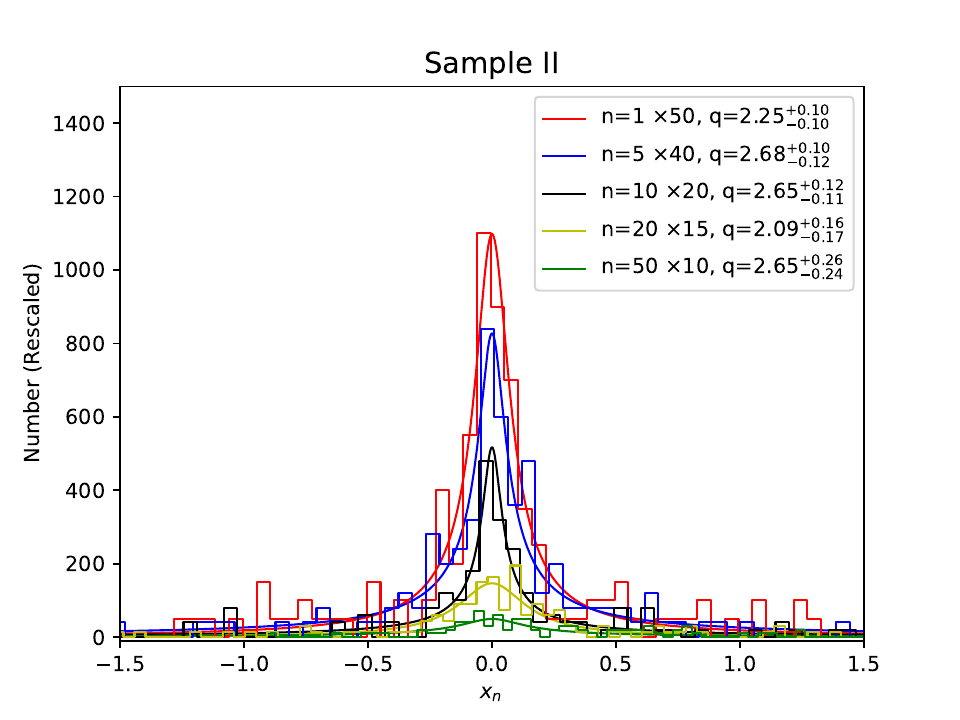}\includegraphics[width=0.5\textwidth, angle=0]{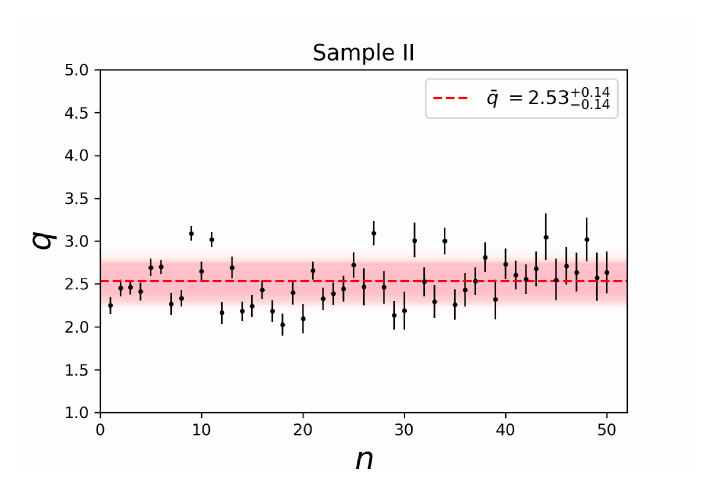}\\
\caption{ Data ($x_n$) distributions and fitting results about $q$-Gauss. The left panel shows several examples of $x_n$ distributions for different values of $n$ and corresponding fitting curves (which have been rescaled to distinguish each sample). The right panel shows the evolution between the best-fitting results of $q$ (with 1-$\sigma$ error) and $n$ of PDFs. }
\label{q-gauss}
\end{figure*}

The Tsallis q-Gaussian distribution of waiting time is also a crucial tool for statistical research. Next, we examine its application to the distribution of $X_{n}$, which is defined as $X_{n}=S_{i+n}-S_{i}$, where $n$ is the temporal interval scale and $S_{i}$ is the scale size of the $i$th waiting time. To reduce computation, we rescaled $X_{n}$ by $\sigma_{X_{n}}$ as
$x_{n}=X_{n}/\sigma_{X_{n}}$, where $\sigma_{X_{n}}$ is the standard deviation of $X_{n}$. The differential distribution of $x_{n}$ can be described with the function given by \cite{1998PhyA..261..534T},
which can be written as
\begin{equation}
f(x_{n})=A\bigl[1-B(1-q)x_{n}^{2}\bigr]^{1/(1-q)}
.\end{equation}
The parameter of interest is $q$, while $A$ and $B$ represent free-fitting parameters.

It is worth mentioning that a theoretical relationship has been established between the power-law index, $\alpha_x$, and the q value of the corresponding q-Gaussian, given by the equation \citep{2010PhRvE..82b1124C} 

\begin{equation}
{\alpha}=\frac{2}{q-1}.
\label{alpha-q}
\end{equation}

\section{Fitting results}
\subsection{Fitting results of the waiting time distribution}

The fitting results of the cumulative distributions of waiting times from Figure \ref{fit-wtd} clearly indicate that the WTD significantly deviates from the exponential distribution. The threshold power-law fitting results of the cumulative distributions suggest that the WTD exhibits a pronounced power-law characteristic.

The fitting results for Sample \uppercase\expandafter{\romannumeral1} and Sample \uppercase\expandafter{\romannumeral2} yield $\alpha_{x}=1.67^{+0.04}_{-0.04}$ and $\alpha_{x}=2.10^{+0.10}_{-0.08}$, respectively, for their cumulative distributions. Here, $\alpha$ represents the power-law index of the threshold power-law functions for each sample.

The fitting results of the distributions show a significant difference between the threshold power law and exponential fits. However, in the fitting of the distributions, the threshold power-law function demonstrates better performance (with $\chi^2_{cum,I}$ = 0.77 and $\chi^2_{cum,II}$ = 1.02).

\subsection{Tsallis q-Gaussian distribution}

After data reduction, it was found that both samples conform to the Tsallis q-Gaussian distribution. The left panel of Figure \ref{q-gauss} shows examples of the probability density function for $n$ = 1, 5, 10, 20, and 50, while the right panel illustrates the evolution of the fitting value of $q$ over the range of $n$ from 1 to 50. The fitting results illustrate that the distributions can be effectively fit by the q-Gaussian function, and the value of $q$ obtained from the fitting remains relatively stable with variations in the value of $n$. The fitting results for Sample \uppercase\expandafter{\romannumeral1} and Sample \uppercase\expandafter{\romannumeral2} show $\bar{q_{\text{I}}}=2.60^{+0.14}_{-0.14}$ and $\bar{q_{\text{II}}}=2.53^{+0.14}_{-0.14}$, respectively.

\section{Discussion }
In terms of the distribution of waiting times, events generated by a Poisson process commonly exhibit an exponential distribution in their waiting times, especially for photons detected by the detector. However, the findings of this study suggest that the WTD of teraelectronvolt photons detected by LHAASO deviates from the exponential distribution and aligns more closely with a power-law distribution. This indicates that these photons are not produced by independent Poisson processes for each event, suggesting a more intricate generation mechanism. \cite{2024MNRAS.533L..64S} propose that the VHE gamma-ray spectra of GRB 221009A observed by WCDA and KM2A at different time intervals and energy levels can be well elucidated using the photo-hadronic model, which includes extragalactic background light models. In this scenario, the generation of VHE photons is attributed to the interaction between high-energy protons and SSC photons in the forward shock region of the jet. \cite{2024arXiv240403229Z} reveal a close relation between the kiloelectronvolt-megaelectronvolt and teraelectronvolt emissions. The prompt emission in the kiloelectronvolt-megaelectronvolt band may continuously inject energy for the generation of teraelectronvolt afterglow. Based on multiwavelength and multimessenger observations, \cite{2023SCPMA..6689511W} discuss the constraints on the properties of GRB ejecta, with the results favoring magnetic field dominance and suggesting the possibility of Poynting flux-dominated GRB ejecta, as well as the potential for strong magnetic field dissipation and acceleration processes.

Based on our power-law fitting results, the SOC theory appears to offer a plausible explanation. According to the theoretical framework proposed by \cite{2012A&A...539A...2A}, we can quantitatively link the concept of fractal dimensions in SOC avalanche systems to the power-law index of system parameters. The theory predicts that the power-law index of the distribution for SOC systems can be defined for Euclidean space dimensions $S=1$, $2$, $3$. \cite{2012A&A...539A...2A} provided some theoretical indices, such as the duration frequency distribution ($\alpha_{T}$), expressed as $\alpha_{T}=\frac{S+1}{2}$, where $S=1$, $2$, and $3$ represent the Euclidean dimensions. Specifically, theoretically, the slopes are $\alpha_{T}=1$, $1.5$, $2$, for $S=1$, $2$, $3$, respectively.
Additionally, some studies have proposed that the generation of teraelectronvolt photons may be associated with Poynting flux-dominated magnetic processes \citep{2023SCPMA..6689511W}, thereby making the existence of a SOC process plausible. The power-law distribution is consistent with a nonstationary process, as was discussed by \cite{1998ApJ...509..448W} and \cite{2010ApJ...717..683A}; for instance, the interaction between external shocks and the circumstellar medium with a time-dependent rate.

\begin{figure}[!http]
\centering
\includegraphics[width=0.5\textwidth, angle=0]{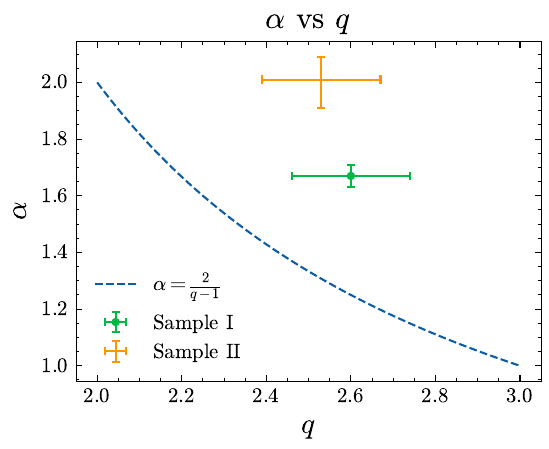}\\
\caption{Comparison between theoretical predictions and fitting results. The dashed blue line in the figure represents the relationship curve  proposed by \cite{2010PhRvE..82b1124C}. The other data points correspond to the fitting results of two samples. }
\label{q-alpha}
\end{figure}

The fitting results of the Tsallis q-Gaussian distribution reveal that the parameter $q$ remains constant with the change of n and exhibits a certain scale invariance. However, this pattern significantly deviates from the theoretical relationship proposed by \cite{2010PhRvE..82b1124C}, as is shown in Fig. (\ref{q-alpha}), which we attribute to a generation mechanism that governs the occurrence of photon wait times, indicating a lack of independence and randomness. This inconsistency can be explained by contradicting the initial assumption in the derivation that the photons generated by the avalanche process are not completely Markovian (independent), especially in Sample \uppercase\expandafter{\romannumeral2} with higher event rates, which results in significant bias at this stage.

\section{Conclusions}
In this work, we investigate the behavior of teraelectronvolt photons from the brightest GRB to date, GRB 221009A. We study the statistical properties of the waiting time of teraelectronvolt photons using various methods. Our main conclusions are as follows:

(1). We find that, dividing the waiting time of teraelectronvolt photons into two samples, a complete one and a partial one, both exhibit well-defined power-law distributions in the cumulative distributions, significantly deviating from exponential distributions.

(2). After fitting the Tsallis $q$-Gaussian distribution to the fluctuations in the data, we observe that the parameter $q$ exhibits a certain degree of constancy with respect to $n$, indicating a scale-invariance structure of avalanche-size differences.

(3). The power-law index, $\alpha_x$, and the scale invariance feature, $q$, exhibit deviations from the theoretical predictions, particularly in Sample \uppercase\expandafter{\romannumeral2}. We believe that this deviation indicates that the appearance of each teraelectronvolt photon is not completely random.

The most important conclusion is that, although power-law distributions and scale invariance features can serve as indicators of SOC, the deviation of the power-law index in Sample \uppercase\expandafter{\romannumeral2} from theoretical predictions suggests that this is not a typical self-organized critical process. The photons are not completely independent of each other, while still exhibiting fundamental characteristics of self-organized critical systems. We believe that this phenomenon may arise due to the energy injection from the central engine influencing the production of teraelectronvolt radiation.

\begin{acknowledgements}
We thank the anonymous referee for thoughtful
comments. We also thank Shao-Lin Xiong, Chen-Wei Wang, Wen-Jun Tan, Wang-Chen Xue and Yan-Qiu Zhang for helpful
discussions. This work is supported by the National Natural Science Foundation of China (Grant Nos. 12494575, 12273009, U2038106, 12373047 and 12494575), the Natural Science Foundation of Jiangxi Province of China (grant No. 20242BAB26012) and the China Manned Spaced Project (CMS-CSST-2021-A12).
\end{acknowledgements}

%
%

\bibliographystyle{aasjournal}
\bibliography{Tev-SOC-bibtex.bib}

\end{document}